\documentclass[10pt,a4paper]{article}

\usepackage[english]{babel}
\usepackage[dvips]{graphicx}
\usepackage{latexsym,amsfonts,amssymb}
\usepackage{amsmath}
\usepackage{eufrak}
\usepackage{epsf}

\usepackage{xypic}
\usepackage{amscd}
\usepackage[all]{xy}
\usepackage{textcomp} 
\usepackage{makeidx} 
\usepackage{mathrsfs} 

\def\eq{\begin{equation}}
\def\en{\end{equation}}

\def\>>{\rangle}
\def\<<{\langle}
\def\tr{{\rm tr}}

\def\sH{\mathscr H}
\def\sT{\mathscr T}

\def\sC{\mathscr C}
\def\sL{\mathscr L}

\def\cC{\mathcal C}

\def\bC{\mathbb C}

\def\Zo{Z^{(0)}}
\def\Ko{K^{(0)}}
\def\Y1{Y^{(1)}}
\def\Ya{Y^{(\alpha)}}

\def\sk{\vskip .4cm}

\begin{document}	



\begin{titlepage}

\vskip .8truecm
\begin{center}
{\Large\bf  Algebraic Bethe Ansatz for the two species ASEP with
\sk
different hopping rates}

\end{center}

\vskip 1truecm
\begin{center}
Luigi Cantini 
\footnote{Universit\'e Paris-Sud, LPTMS, UMR8626,  B\^at. 100, Universit\'e
  Paris-Sud 91405 Orsay cedex.}
\footnote{CNRS, LPTMS, UMR8626, B\^at. 100, Universit\'e Paris-Sud 91405 Orsay
  cedex.}
\vskip 3.8truecm

\begin{abstract}

An ASEP with two species of particles and different hopping rates is
considered on a ring. Its integrability is proved and the Nested Algebraic Bethe
Ansatz is used to derive the Bethe Equations for states with arbitrary
numbers of particles of each type, generalizing  the results of
Derrida and Evans \cite{derr-evans}. We present also formulas for the total
velocity of particles of a given type and their limit for large size
of the system and finite densities of the particles.

\end{abstract}
\bigskip

%
%

\vskip 1truecm

\end{center}
\end{titlepage}

\section{Introduction}

An idea which has proved to be quite useful in understanding the behavior of
systems out of equilibrium is to study solvable models. An example of such
models, which has
driven a lot of attention for at last two decades, is the asymmetric simple
exclusion process (ASEP) \cite{liggett}. It describes a driven lattice gas
\cite{KLS, SZ}
where particles can hop on adjacent sites with asymmetric rates and hard
core exclusion.   

Different methods have been applied to study the ASEP, and each of them seems 
better suited to study certain aspects of the problem. The Matrix Product
Ansatz for example has been employed with success  in determining 
the density profile of steady states, steady currents or diffusion
coefficients (for reviews see
\cite{derr-rev1, schutz, evans-rev}). On the other hand other quantities
like the relaxation time 
are more easily dealt with by means of the Bethe Ansatz \cite{dhar, gwa-spohn}. 

Actually in \cite{der-lebo1},
Derrida and Lebowitz showed how a modification of the Bethe Ansatz of Gwa
and Sphon \cite{gwa-spohn}, could be used to compute the full large
deviation function 
of the time averaged current for the ASEP with one specie of particles. 

Shortly later Derrida and Evans \cite{derr-evans} considered the problem with a second
specie of particles, and thanks to a Bethe Ansatz they were able, not only to
reproduce known results \cite{derrida96, mallick96}  about the phase diagram of the steady
current of a particle of second 
type, but also to compute its diffusion coefficient and in principles all
the higher cumulants. 

Multi-species generalizations of the ASEP have been considered in several
papers \cite{DJLS,EGFM,AHR,ADR,karimipour}.
The fact that they are integrable is not at all obvious. The
most natural integrable generalizations of the single specie have a
hierarchical structure based on quotients of 
the Hecke algebra, which ensure the
integrability, as explained in \cite{rittenberg}. 

On the other hand, in the ASEP with two kind of
particles and different rates, as considered by Derrida and Evans, the
hierarchy is partially spoiled precisely 
by the different hopping rates, and to understand its integrability from a
point of view of the Yang-Baxter equation one cannot make resort to the Hecke
algebra commutation relations. 

Our first point in the present paper is to make manifest the integrability
of the ASEP with two species, by showing an $R-$matrix which solve the
Yang-Baxter equation and gives, through the usual procedure, the transition
matrix of our problem. Once we have this we employ the machinery of the
Algebraic Bethe Ansatz (ABA) to derive the Bethe equations and the
eigenvalues of the transition matrix (see \cite{faddeev} for a review on
ABA, and \cite{golinelli-mallick1} for recent application of ABA to the ASEP).

Since we have a number of species greater than one we are led to perform a
Nested Bethe Ansatz. In the case of arbitrary number of species, but with
hopping rates independent of the types of particles the Nested Bethe
Equations have already been derived in \cite{Bariev}. 

With the Bethe Equations at disposal we can tackle the problem of determining
the cumulants of the total velocity of particles of a given type,
or of joint cumulants, in presence of an arbitrary number of particles of
each kind. We present the exact formula for the
average velocity of particles of second type and consider as well the limit
of large size of the system with finite non-zero densities of particles. We
comment also on the difficulties about the determination of the higher
cumulants.

The plan of the paper is the following. In section $2$ we show the
integrability of the ASEP with generic rates by presenting an
$R-$matrix which solves the Yang-Baxter equation and generates the
transition matrix of the ASEP. In the same section we use the
techniques of the Algebraic Bethe Ansatz to diagonalize the transition
matrix, arriving at a set of Nested Bethe equations. In section $3$ we
analyze the Bethe equations and derive the exact formula for the total
velocity of $M_2$ particles of type $2$ on a ring of size $N$ and in
presence of $M_1$ particles of kind $1$, we comment on the derivation of the
higher cumulants. The large $N$ limit of the
velocity is worked out in section $4$ where we show that for non-zero
densities of particles of each type, there is no phase transition.  
In Appendix A we sketch the derivation of the Bethe Equations for an ASEP
with twisted boundary conditions, that we need for the derivation of the
Nested Equations. In Appendix B we study the integrability of models with
higher number of particles and arbitrary hopping rates.

\section{Yang-Baxter for two species and different rates}

In \cite{derr-evans}, Derrida and Evans have employed the coordinate  Bethe
Ansatz to 
study an ASEP in presence of an impurity, which in the following will be
treated  as
a second specie of particles. If we indicate with $0$ the empty
site (which can be considered as a particle of type zero), $1$ a
particle of first kind and $2$ a particle of second kind, 
then the rules that govern the
stochastic evolution of the system during an interval of time $dt$ are
purely local on couples of neighboring sites and are given by 
\begin{center}
$10 \rightarrow 01$ with rate $1$

$20 \rightarrow 02$ with rate $\alpha$

$12 \rightarrow 21$ with rate $\beta$.
\end{center}

The fact that the problem can be solved by Bethe Ansatz, as done in
\cite{derr-evans} for the case of a single second type particle,  means
that it is integrable. Our first task is to understand better its 
integrability showing the Yang-Baxter equation behind it.  


The transition matrix of our system can be written in terms of matrices
that encode the local transition of particles. Let us define the basis
of the local space of states as: 
\begin{equation*}
\begin{split}
|0\>> &= \textrm{empty} \equiv \textrm{particle of kind}~~ 0,\\ 
|1\>> &= \textrm{particle of kind}~~ 1,\\ 
|2\>> &= \textrm{particle of kind}~~ 2. 
\end{split}
\end{equation*}
In the basis $(|0\>>, |1\>>, |2\>>)_a \otimes (|0\>>, |1\>>,
|2\>>)_b$ we introduce the matrices
$$
E^{(10)} = 
\left(
\begin{array}{ccccccccc}
0 &0 &0 &0 &0 &0 &0 &0 &0\\
0 &0 &0 &e^{\nu_{10}} &0 &0 &0 &0 &0\\
0 &0 &0 &0 &0 &0 &0 &0 &0\\
0 &0 &0 &-1 &0 &0 &0 &0 &0\\
0 &0 &0 &0 &0 &0 &0 &0 &0\\
0 &0 &0 &0 &0 &0 &0 &0 &0\\
0 &0 &0 &0 &0 &0 &0 &0 &0\\
0 &0 &0 &0 &0 &0 &0 &0 &0\\
0 &0 &0 &0 &0 &0 &0 &0 &0
\end{array}
\right);~~~
E^{(20)} = 
\left(
\begin{array}{ccccccccc}
0 &0 &0 &0 &0 &0 &0 &0 &0\\
0 &0 &0 &0 &0 &0 &0 &0 &0\\
0 &0 &0 &0 &0 &0 &e^{\nu_{20}} &0 &0\\
0 &0 &0 &0 &0 &0 &0 &0 &0\\
0 &0 &0 &0 &0 &0 &0 &0 &0\\
0 &0 &0 &0 &0 &0 &0 &0 &0\\
0 &0 &0 &0 &0 &0 &-1 &0 &0\\
0 &0 &0 &0 &0 &0 &0 &0 &0\\
0 &0 &0 &0 &0 &0 &0 &0 &0
\end{array}
\right);~~~
$$

\eq
E^{(12)} = 
\left(
\begin{array}{ccccccccc}
0 &0 &0 &0 &0 &0 &0 &0 &0\\
0 &0 &0 &0 &0 &0 &0 &0 &0\\
0 &0 &0 &0 &0 &0 &0 &0 &0\\
0 &0 &0 &0 &0 &0 &0 &0 &0\\
0 &0 &0 &0 &0 &0 &0 &0 &0\\
0 &0 &0 &0 &0 &-1 &0 &0 &0\\
0 &0 &0 &0 &0 &0 &0 &0 &0\\
0 &0 &0 &0 &0 &e^{\nu_{12}} &0 &0 &0\\
0 &0 &0 &0 &0 &0 &0 &0 &0
\end{array}
\right);~~~
\en
The ASEP with two species is defined by the following equation for the
probability of a given configuration $\cC$
\eq\label{transition2}
\frac{d}{dt}P_t(\cC) = \sum_{\cC'} M^{0,0,0}(\cC, \cC')P_t(\cC'). 
\en
where 
\eq
M^{\nu_{10}, \nu_{20}, \nu_{12}}(\cC, \cC') = \sum_i (E^{(10)}_i + \alpha E^{(20)}_i +
\beta E^{(12)}_i). 
\en

Since we are in presence of several species of particles we can introduce the
\emph{relative} distances 
covered by particles after an initial time $t=0$. By this we mean  the
distance $Y^{ij}$ covered  
by all the particles of kind $i$ with respect to particles of kind
$j$. It increases of a unity each time a particle of kind $i$ jumps
to the right of a particle of kind $j$, and decreases of a unity when
the opposite happens. In our case we have three kind of particles
$0, 1, 2$, hence we consider $Y^{10}$, $Y^{20}$, $Y^{12}$. The
joint probability $P_t(\cC,Y^{10}, Y^{20}, Y^{12})$ of being in a
configuration $\cC$, and having $Y^{ij}_t = Y^{ij}$ satisfies an
evolution equation which is better written in terms of a generating
function
\eq
F^{\nu_{10}, \nu_{20}, \nu_{12}}_t(\cC) = \sum_{Y^{10}, Y^{20}, Y^{12}}
e^{\nu_{10} Y^{10} + \nu_{20} Y^{20} +\nu_{12} Y^{12}}P_t(\cC,Y^{10}, Y^{20}, Y^{12}).
\en  
The evolution equation satisfied by $F^{\nu_{10}, \nu_{20}, \nu_{12}}_t(\cC)$ is
\eq\label{transition3}
\frac{d}{dt}F^{\nu_{10}, \nu_{20}, \nu_{12}}_t(\cC) = \sum_{\cC'} M^{\nu_{10},
  \nu_{20}, \nu_{12}}(\cC, \cC')F^{\nu_{10}, \nu_{20}, \nu_{12}}_t(\cC').  
\en
One obtains $\<<
e^{\nu_{10} Y_t^{10} + \nu_{20} Y_t^{20} +\nu_{12} Y_t^{12}}\>>$ summing
$F^{\nu_{10}, \nu_{20}, \nu_{12}}_t(\cC)$ over $\cC$ , hence its 
large time behavior 
is determined
by the largest 
eigenvalue  $\lambda(\nu_{10}, \nu_{20}, \nu_{12})$ of  the transition matrix
$M^{\nu_{10},  \nu_{20}, \nu_{12}}(\cC, \cC')$
$$
\<<
e^{\nu_{10} Y_t^{10} + \nu_{20} Y_t^{20} +\nu_{12} Y_t^{12}}\>>  \sim
e^{\lambda(\nu_{10}, \nu_{20}, \nu_{12}) t}.
$$ 
We find such an eigenvalue  by employing the Algebraic Bethe
Ansatz. 

Our first step is to find an $R-$matrix which satisfies the Yang-Baxter equation, the
inversion relation and
such that its derivative reduces to the linear combination of $E^{(ij)}$'s
matrices
$$
E^{(10)} + \alpha E^{(20)}  + \beta E^{(12)}
$$
 
Once we have this we construct the transfer matrix in the usual
way as trace of product of $L-$matrices ($L= P R$) and we are insured that
its logarithmic derivative will be the desired
$\sum_i (E^{(10)}_i + \alpha E^{(20)}_i + \beta E^{(12)}_i)$
We provide a solution of the Yang-Baxter equation 
\eq\label{YB}
R_{a,b}(y,z) R_{b,c}(x,z) R_{a,b}(x,y) = R_{b,c}(x,y) R_{a,b}(x,z) R_{b,c}(y,z).
\en
of the form
\eq\label{baxterized}
R(x,y) = 1 + g_{10}(x,y) E^{(10)} + g_{20}(x,y) E^{(20)}  + g_{12}(x,y) E^{(12)},
\en
where 
\eq
g_{12}(x,y) = 1 - \frac{1+\beta(e^{-y}-1)}{1+\beta(e^{-x}-1)}; ~~~ g_{10}(x,y) = 1 - e^{x-y};
~~~ g_{20}(x,y) = 1 - \frac{1+\alpha(e^{x}-1)}{1 + \alpha(e^{y}-1)}.
\en
We define the monodromy matrix of a system of size $N$ as 
\eq
\sT_{a\otimes \sH}(x,\vec \eta) = L_{a, a_N}(x,\eta_N)\dots
L_{a,a_2}(x,\eta_2) L_{a,a_1}(x,\eta_1). 
\en
where $L_{a,b}(x,y)= P_{a,b}  R_{a,b}(x,y)$, and $P_{a,b}$ is the
permutation operator, i.e. $P v_a\otimes v_b = v_b\otimes v_a$. .
The transfer matrix is given by
$$
T(x,\vec \eta) = \tr_a\sT_{a\otimes \sH}(a,\vec \eta)
$$
Thanks to the Yang-Baxter equation (\ref{YB}) we get
\eq\label{commutation3}
\sT_{a\otimes \sH}(z,\vec y)  \sT_{b\otimes \sH}(x,\vec y) R_{a,b}(x,z)
= R_{a,b}(x,z) \sT_{a\otimes \sH}(x,\vec y) \sT_{b\otimes \sH}(z,\vec y).
\en
and, tracing over the auxiliary space, we obtain that the transfer matrices
with different values of the spectral parameter commute among themselves 
\eq
[T(x,\vec \eta),T(x',\vec \eta)]=0.
\en
The transition matrix of our system is obtained choosing $\eta_i=0$,
and taking the logarithmic derivative of $T(x,\vec \eta)$ at $x=0$
\eq\label{hamilton}
M^{\nu_{10},  \nu_{20}, \nu_{12}}(\cC, \cC') =  -T(0,\vec 0)^{-1}\frac{d
  T(x,\vec 0) }{dx}\Big|_{x=0}. 
\en

\sk

We come now to the Yang-Baxter algebra, which can be easily read from
eq.(\ref{commutation3}). Let us write the monodromy matrix as 
\eq
\sT_{a\otimes \sH}(x) = \left(
\begin{array}{ccc}
A(x) & B_1(x) & B_2(x)\\
C_1(x) & D_{11}(x) & D_{12}(x)\\
C_2(x) & D_{21}(x) & D_{22}(x)
\end{array}
\right)
\en
The transfer matrix can then be written as 
\eq\label{transf-A+D}
T(x) = A(x) + D_{11}(x) + D_{22}(x),
\en
where we have simplified the notation omitting $\vec \zeta$ which is fixed
to be zero.
For later purposes let us rewrite the $R-$ matrix as
\eq
R(x,y) = 
\left(
\begin{array}{ccccccccc}
1 & 0 &0 &0 &0 &0 &0 &0 &0 \\
0 & 1 & 0& (1-e^{x-y})e^{\nu_{10}}  &0 &0 &0 &0 &0 \\
0 &0 &1 &0 &0 &0 &(1-\frac{1 + \alpha(e^x-1)}{1 + \alpha(e^y-1)})e^{\nu_{20}} & & \\
0 &0 &0 &e^{x-y} &0 &0 &0 &0 &0 \\
0 &0 &0 &0 &R^{(1)}_{11,11} &R^{(1)}_{11,12} &0
&R^{(1)}_{11,21}  & R^{(1)}_{11,22} \\
0 &0 &0 &0 &R^{(1)}_{12,11} &R^{(1)}_{12,12} &0
&R^{(1)}_{12,21}  & R^{(1)}_{12,22} \\
0 &0 &0 &0 &0 &0 &\frac{1 + \alpha(e^x-1)}{1 + \alpha(e^y-1)} &0 &0 \\
0 &0 &0 &0 &R^{(1)}_{21,11} &R^{(1)}_{21,12} &0
&R^{(1)}_{21,21}  & R^{(1)}_{21,22} \\
0 &0 &0 &0 &R^{(1)}_{22,11} &R^{(1)}_{22,12} &0
&R^{(1)}_{22,21}  & R^{(1)}_{22,22} 
\end{array}
\right)
\en
The matrix $R^{(1)}$, which actually depends on $x$ and $y$, is given
by 
\eq
R^{(1)}(x,y) =\left(
\begin{array}{cccc}
1&0&0&0\\
0&\frac{1+\beta(e^{-y}-1)}{1+\beta(e^{-x}-1)}&0&0\\
0&(1-\frac{1+\beta(e^{-y}-1)}{1+\beta(e^{-x}-1)})e^{\nu_{12}}&1&0\\
0&0&0&1
\end{array}
\right).
\en
It is nothing else than the $R-$matrix corresponding to the ASEP with a
single specie presented in eq.(\ref{R1}) in appendix A, with a different
parameterization
\eq\label{param}
e^{y_i} \rightarrow \frac{1}{1+\beta(e^{-y_i}-1)}.
\en

With this notation we can write the commutation rules of the operators $A,
B_i, C_i$ and $D_{ij}$ appearing in $\sT$, for different values of the
spectral parameters
\begin{eqnarray}
&[A(x),A(y)] = 0; \\
&A(x)B_1(y) = \frac{e^x}{e^{\nu_{10}}(e^x-e^y)}B_1(y)A(x) +
\frac{e^y}{e^{\nu_{10}}(e^y-e^x)}B_1(x)A(y);\\
&A(x)B_2(y) = \frac{1+ \alpha(e^x-1)}{e^{\nu_{20}}\alpha(e^x-e^y)}B_2(y)A(x) +
\frac{1+\alpha(e^y-1)}{e^{\nu_{20}}\alpha(e^y-e^x)}B_2(x)A(y);\\  
&B_i(x)B_j(y) = B_l(y) B_k(x) R^{(1)}_{i j, l k}(x,y);\\
&D_{1j}(x)B_k(y) = \frac{e^y}{e^{\nu_{10}}(e^y-e^x)}(B_m (y) D_{1
  n}(x)R^{(1)}_{j k, m n}(x,y) - B_j(x) D_{1 k}(y))  ;\\
&D_{2j}(x)B_k(y) = \frac{1+\alpha(e^y-1)}{e^{\nu_{20}}\alpha(e^y-e^x)}(B_m (y) D_{2 
  n}(x){R^{(1)}}_{j k, m n}(x,y) - B_j(x) D_{2 k}(y)).
\end{eqnarray}

The ansatz for an eigenvector of the transfer matrix, keeping into
account the non commutativity of 
the $B_i$s for different $i$, is given by
\eq\label{ansatz11}
|\Psi^{M_1,M_2}(y_1,\dots,y_{r})\>> = \sum_{i_1,\dots, i_r}
\Psi^{M_1,M_2}_{i_1,\dots, i_r} B_{i_1}(y_1) \dots B_{i_r}(y_r) ||1\>>,
\en
where $||1\>>$ is the reference state, defined by
$$
||1\>>= 
\left( \begin{array}{c}
1\\
0\\
0
\end{array}  \right)
\otimes
\dots
\otimes
\left( \begin{array}{c}
1\\
0\\
0
\end{array}  \right),
$$
and it is an eigenstate separately of $A(x), D_{11}(x)$
and $D_{22}(x)$
$$
A(x) ||1\>> = ||1\>>, ~~~~D_{11}(x) ||1\>>= e^{N\nu_{10}}
(1-e^{x})^N||1\>>, ~~~~ 
$$
$$
D_{22}(x)||1\>>
=(\alpha e^{\nu_{20}})^N (1-e^{x})^N ||1\>>,  
$$
and correspond to a completely empty system.
The labels $M_1, M_2$ in eq.(\ref{ansatz11}) mean that we require $M_1$ $B$s of type $1$
and $M_2$ $B$s of type $2$, i.e. we are restricting to the sector with $M_1$
particles of type $1$ and $M_2$ particles of type $2$.

The eigenvector equation for $|\Psi^{M_1,M_2}(y_1,\dots,y_{r})\>>$ puts
constraints on the $y_i$s.
Let us first apply the operator $A(x)$ to
$|\Psi^{M_1,M_2}(y_1,\dots,y_{r})\>>$. We get a wanted term, i.e. a term
proportional to the vector we start from, of 
 the form 
\eq\label{wantedA}
\left(\frac{e^x}{e^{\nu_{10}}}\right)^{M_1} \left(\frac{1+\alpha (e^x-1)}{e^{\nu_{20}}\alpha} \right)^{M_2}
\prod_{i=1}^{M_1+M_2} 
\frac{1}{(e^x-e^{y_i})} |\Psi^{M_1,M_2}(y_1,\dots,y_{r})\>>
\en
and unwanted terms of the form
$$
\frac{1}{(e^{y_j}-e^x)}\left(\frac{e^{y_j}}{e^{\nu_{10}}}\right)^{M_1} \left(\frac{1+\alpha
    (e^{y_j}-1)}{e^{\nu_{20}}\alpha}\right)^{M_2} 
\prod_{i\neq j}^{M_1+M_2}
\frac{1}{(e^{y_j}-e^{y_i})} B(x)\otimes B(y_{j+1})\otimes \dots
\otimes B(y_{j-1}) 
$$
\eq
M(y_j,\vec y)M(y_{j-1},\vec y) \dots
M(y_1,\vec y)\Psi^{M_1,M_2}(y_1,\dots,y_{r}) ||1\>>
\en
with 
$$
M(y_j,\vec y) = R^{(1)}_{i_1 i_2,j_1 j'_2}(y_j,y_{j+1})
R^{(1)}_{j'_2 i_3,j_2 j'_3}(y_j, y_{j+2})\dots  R^{(1)}_{j'_{r-1}
  i_r,j_{r-1},j_r}(y_j, y_{j-1})
$$
The unwanted terms have to cancel with similar terms coming from the action
of 
$D_{11}(x,\vec y)+ D_{22}(x,\vec y)$.
From the action of $D_{kk}$ we get a wanted term
\eq\label{wantedD}
\omega_k(\vec y)(1-e^x)^N \prod_{i=1}^{M_1+M_2}\frac{1}{(e^{y_i}- e^x)}
B(y_1)\otimes \dots  
\otimes B(y_{M_1+M_2})T^{(1)}_{kk}(x,\vec y) \Psi^{M_1,M_2}(y_1,\dots,y_{r})
||1\>>, 
\en
with 
$$
\omega_1(\vec y) =
e^{N\nu_{10}}\prod_{i=1}^{M_1+M_2}\left(\frac{e^{y_i}}{e^{\nu_{10}}}\right),~~~~~~\omega_2(\vec
y) = e^{N\nu_{20}}\alpha^{N} 
\prod_{i=1}^{M_1+M_2} \left(\frac{1+\alpha(e^{y_i}-1)}{e^{\nu_{20}}\alpha}\right),
$$
and
\eq
T^{(1)}(x,\vec y) = L^{(1)}_{a,a_{M_1+M_2}}(x, y_{M_1+M_2})\dots
  L^{(1)}_{a,a_1}(x, y_1)  
\en
is just the monodromy matrix of TASEP with a single species as explained in
the appendix \ref{ABA1}.
We get also an unwanted term
$$
\omega_k(\vec y)\frac{(1-e^{y_j})^N}{e^x -e^{y_j}} \prod_{i \neq j}^{M_1+M_2}\frac{1}{(e^{y_i}- e^{y_j})}
 B(x)\otimes B(y_{j+1})\otimes \dots
\otimes B(y_{j-1})\times 
$$
$$
M(y_j,\vec y)M(y_{j-1},\vec y) \dots
M(y_1,\vec y)\Psi^{M_1,M_2}(y_1,\dots,y_{r}) T^{(1)}_{kk}(y_j,\vec y)\Psi^{M_1,M_2}(y_1,\dots,y_{r})
||1\>>. 
$$
In order to get the cancellation of the unwanted terms we first have to
diagonalize $\omega_1(\vec y) T^{(1)}_{11} + \omega_2(\vec y)
T^{(1)}_{22}$. This is the transfer matrix of a TASEP with a single specie
and twisted boundary condition and can be diagonalized by the Algebraic Bethe
Ansatz as in the case of non-twisted boundary conditions, as done in
\cite{golinelli-mallick1}, we briefly recall how it works in appendix
\ref{ABA1}. Here 
we simply apply  
the results explained there. 
One has only to be careful in 
translating the parameters $e^{\tilde y_i}$ appearing in the appendix,
following eq.(\ref{param}) 
\eq
e^{\tilde y_i} = \frac{1}{1+\beta(e^{-y_i}-1)} = \frac{Y_i-1}{(1-\beta)Y_i-1},
\en 
where we have defined $Y_i = 1 - e^{y_i} $, and remember that we consider only the sector
with $M_2$ particles in a system of size $\tilde N = M_1+M_2$.

For the auxiliary spectral parameters $Z$s we get the Bethe equation
(\ref{bethe-1sp}) which now reads
\eq\label{bethe1}
\left(\frac{e^{\nu_{20}} \alpha}{e^{\nu_{10}}}\right)^N \prod_{i=1}^{M_1+M_2}\frac{(1-\alpha
  Y_i)(b Y_i
  -1)}{(1-Y_i)(b Y_i-1 - Z_j (Y_i-1))}\prod_{k\neq j}^{M_2} \left(-
  \frac{Z_j}{Z_k} \right) = \left(\alpha \frac{e^{\nu_{12}} e^{\nu_{20}}}{e^{\nu_{10}}}\right)^{M_1+M_2}. 
\en
While the cancellation of the unwanted terms coming from $A$ and $D_{kk}$
leads to a second Bethe equation
\eq\label{bethe2}
\frac{(1-Y_j)^{M_1}(1-\alpha Y_j)^{M_2}(b Y_j-1)^{M_2}}{e^{N\nu_{10}}Y_j^N 
\prod_{i=1}^{M_1+M_2} \left( 1- Y_i  \right)} = \left ( \frac{\alpha e^{\nu_{12}} 
e^{\nu_{20}}}{e^{\nu_{10}}}\right )^{M_2} 
\prod_{k=1}^{M_2} \left (b Y_j-1 - Z_k (Y_j-1)\right).   
\en
The eigenvalue of the transfer matrix can be read from eqs.(\ref{wantedA},
\ref{wantedD}) and eq.(\ref{eigen1})
\eq\label{lambda}
\Lambda(x) = \left(\frac{e^x}{e^{\nu_{10}}}\right)^{M_1} \left(\frac{1+\alpha
    (e^x-1)}{e^{\nu_{20}}\alpha} \right)^{M_2} \prod_{i=1}^{M_1+M_2} 
\frac{1}{(e^x-e^{y_i})} + (1-e^x)^N e^{\nu_{12} M_2} \omega_1(\vec y) \prod_{i=1}^r
\left(1-e^{x}Z_i \right)\prod_{i=1}^{M_1+M_2} 
\frac{1}{(e^x-e^{y_i})}. 
\en
Taking the logarithmic derivative in $x=0$ we get the eigenvalue of the
transition matrix $\lambda$
\eq\label{lambda}
\lambda = -\frac{1}{\Lambda(0)}\frac{d \Lambda(x)}{dx}\Big| _0 = -M_1 - \alpha M_2 +
\sum_{i=1}^{M_1+M_2} \frac{1}{Y_i}.
\en

\section{Analysis of the Bethe equations}

For convenience of notation we divide the $Y$s in two sets, $\Y1_i$ with
$i=1,\dots 
M_1$, and $\Ya_i$ with $i=1,\dots M_2$. The solution of the Bethe
equations wanted behaves in the limit 
$\nu_{ij} \rightarrow 0$ as $Y_i^{(1)} \rightarrow 1$, $Y_i^{(\alpha)}
\rightarrow 1/\alpha$ and $Z_k \rightarrow Z_k^{(0)}$, where the
$Z_k^{(0)}$ have to be determined. Actually we will see that the
$Z_k^{(0)}$s depend on how the limit is taken. For this reason we will
redefine $\nu_{ij} \rightarrow \nu \nu_{ij}$ and take the limit $\nu
\rightarrow 0$ keeping $\nu_{ij}$ fixed. What happens is that the
$Z_k^{(0)}$s depend on these $\nu_{ij}$ (or more precisely on their
ratios).

From eqs.(\ref{bethe1},\ref{bethe2}) we get \mbox{$(e^{\nu_{20} M_2 +
    \nu_{10}M_1}
  \alpha^{M_2}\prod_{i=1}^{M_1}Y_i^{(1)}\prod_{i=1}^{M_2}Y_i^{(\alpha)})^N=1$} 
and by continuity 
\eq\label{product}
e^{\nu_{20} M_2 + \nu_{10}M_1}
\alpha^{M_2}\prod_{i=1}^{M_1}Y_i^{(1)}\prod_{i=1}^{M_2}Y_i^{(\alpha)}~=~1
\en
Let us introduce the following auxiliary variables
\eq\label{def2}
C =  e^{(\nu_{12}+\nu_{20}-\nu_{10})M_2 + \nu_{10}N} \prod_{i=1}^{M_1+M_2} (1-Y_i),~~~~~~ K =
-\frac{e^{\nu_{12} (M_1+M_2)}}{(\alpha e^{\nu_{20}- \nu_{10}})^{N-M_1-M_2}}\frac{\prod_{i=1}^{M_1+M_2}
  (1-Y_i)~\prod_{i=1}^{M_2} Z_k}{\prod_{i=1}^{M_1+M_2} (1 - \alpha   
Y_i)(bY_i-1)}, ~~~~~~
\en
The Bethe equations become 
\eq\label{B-1}
(-Z_j)^{M_2} = K \prod_{i=1}^{M_1+M_2}(b Y_i-1 - Z_j (Y_i-1))
\en
\eq\label{B-2}
(1-Y_j)^{M_1}(1-\alpha Y_j)^{M_2}(b Y_j-1)^{M_2}= C \alpha^{M_2} Y_j^N 
\prod_{k=1}^{M_2} \left (b Y_j-1 - Z_k (Y_j-1)\right).  
\en
We notice that if we keep $K$ as an unknown, combining eq.(\ref{product})
with eqs.(\ref{B-1}, \ref{B-2}) we recover the definition of $K$ given in
eq.(\ref{def2}). Hence from now on our basic equations are (\ref{product}),
(\ref{B-1}) and (\ref{B-2}).
Following steps similar to the ones in \cite{derr-evans} we obtain the
following representation for the eigenvalue, which generalizes eqs. ($33$, $34$
and $36$) of
\cite{derr-evans} 
\eq\label{lambda}
\lambda = - \sum_{n=1}^\infty \frac{C^n}{n}\left[\oint_1 +
  \oint_{1/\alpha} \right] \frac{dy}{2\pi i } \frac{1}{y^2}[Q(y)]^n
\en
where for us 
\eq
Q(y) = \frac{y^N \alpha ^{M_2} \prod_{k=1}^{M_2}(by -1 -Z_k(y-1))}{(1-\alpha
  y)^{M_2}(by-1)^{M_2}(1-y)^{M_1}}
\en
Taking the logarithm of eq.(\ref{product}) we get 
\eq\label{nu}
\nu_{10} M_1 + \nu_{20} M_2 =  - \sum_{n=1}^\infty \frac{C^n}{n}\left[\oint_1 +
  \oint_{1/\alpha} \right] \frac{dy}{2\pi i } \frac{1}{y}[Q(y)]^n,
\en
while eq.(\ref{B-1}) becomes (after having taken the logarithm)
\eq\label{eq-Z}
M_2 \log (-Z_j) = \log K + 2i\pi j + M_1 \log(b-1) + M_2 \log(b/\alpha -1 -
Z_j(1/\alpha -1)) + 
\en
$$
+\sum_{n=1}^{\infty}\frac{C^n}{n}\left[\oint_1 +
  \oint_{1/\alpha} \right]\frac{dy}{2\pi
  i}\frac{b-Z_j}{by-1-Z_j(y-1)}Q(y)^n. 
$$
Taking the logarithm of the first equation in (\ref{def2}) we get also the
equation 
\eq\label{constr1}
\nu_{10} (N-M_2) + (\nu_{20}+ \nu_{12}) M_2  =  - \sum_{n=1}^\infty \frac{C^n}{n}\left[\oint_1 +
  \oint_{1/\alpha} \right] \frac{dy}{2\pi i } \frac{1}{y-1}[Q(y)]^n.
\en

Now, if we redefine $\nu_{ij} \rightarrow \nu \nu_{ij}$, we see that
eqs.(\ref{lambda}, 
\ref{nu}, \ref{eq-Z} and \ref{constr1}) give in an implicit form the power
expansion in $\nu$ of $\lambda$. What one should do in principle is to expand $
log(K)$ and $Z_j$ in powers of $C$, use eq. (\ref{eq-Z}) and a
combination of  eqs.(\ref{nu},\ref{constr1}) to derive the n-th order term
of these expansions in terms of lower orders terms, and then work out the
expansion of  $\lambda$ in powers of $\nu$. This way one would find the
cumulants of the total number of particles flown  $\nu_{10}
Y^{(10)}+\nu_{20} Y^{(20)}+\nu_{12} Y^{(12)}$  
\eq
\lambda(\nu) = \lim_{t\rightarrow \infty} \frac{\log \<<e^{\nu(\nu_{10}
Y^{(10)}+\nu_{20} Y^{(20)}+\nu_{12} Y^{(12)})}\>>}{t} = \sum_n \frac{\<< (\nu_{10}
Y^{(10)}+\nu_{20} Y^{(20)}+\nu_{12} Y^{(12)})^n\>>_c}{t} \nu^n.
\en

Concretely this is of course quite
laborious and one doesn't find any illuminating formulas in general, but
one can easily find at least the velocities. Let us work out explicitly the
average of the total velocity of the particles of second type. For this we
have to chose $\nu_{10}=0$, $\nu_{20} = -\nu_{12} =1$, and the velocity is
given simply by the linear term of the
expansion of $\lambda$ in terms on $\nu$.  
In the limit $\nu \rightarrow 0$, the $Z_j$s satisfy a very simple equation 
\eq
(\Zo_j)^{M_2} = (-1)^{(M_2)} \Ko (b-1)^{M_1}\left[\frac{b}{\alpha}-1 - \Zo_j
\left(\frac{1}{\alpha} -1  \right)\right]^{M_2},
\en
whose solution is
\eq
\Zo_j = \frac{(\alpha-b)e^{\frac{2\pi i j}{M_2}}[\Ko(b-1)^{M_1}]^{
    1/M_2}}{\alpha -(1-\alpha)e^{\frac{2\pi i j}{M_2}}[\Ko(b-1)^{M_1}]^{
    1/M_2}}.
\en
The $\Zo_j$ are now expressed in terms of a single unknown $\Ko$ which is
determined taking the first order in $C$ of the constraint equation
(\ref{constr1})
\eq
\left[ \oint_1 + \oint_{1/\alpha} \right] \frac{dy}{2\pi i(y-1)} Q^{(0)}(y)
= 0,
\en 
where $Q^{(0)}$ is the value of $Q$ for $\nu=0$, which is given by
\eq\label{Q0}
Q^{(0)}(y) = \frac{y^N}{(1-y)^{M_1}}\left( \frac{\alpha^{M_2}}{(1-\alpha
    y)^{M_2}} -\Ko \frac{(b-1)^{M_1+M_2}}{(by -1)^{M_2}} \right).
\en
Then for $\Ko$ we find
\eq
\Ko = \frac{\left[ \oint_1 + \oint_{1/\alpha} \right] \frac{dy}{2\pi
    i}  \frac{y^N \alpha^{M_2}}{(1-y)^{M_1+1}(1-\alpha y)^{M_2}}}
{\left[ \oint_1 + \oint_{1/\alpha} \right] \frac{dy}{2\pi i}\frac{y^N
    (b-1)^{M_1+M_2}}{(1-y)^{M_1+1}(b y -1)^{M_2}}}.
\en
Notice that, as stated before, the value of $\Ko$, and hence of the
$\zeta_k^{(0)}$s, depends on the choice of $\nu_{ij}$. Had we chosen
different values for $\nu_{10}$, $\nu_{20}$ and $\nu_{12}$, we would have
found a different value for $\Ko$. 

Now we have all the ingredient we need to derive the velocity of the 
particles of 
kind $2$. We have simply to consider the linear part of eqs.(\ref{lambda},
\ref{nu}) 
\eq
v_2= \lim_{\nu \rightarrow 0} \frac{\lambda(\nu)}{\nu} = M_2 \frac{\left[
    \oint_1 + \oint_{1/\alpha} \right] \frac{dy}{2\pi 
    i } \frac{Q^{(0)}(y)}{y^2} }{\left[ \oint_1 + \oint_{1/\alpha} \right] \frac{dy}{2\pi
    i } \frac{Q^{(0)}(y)}{y} },
\en
which is more conveniently written in terms of the following two auxiliary
functions 
\eq
F^\alpha_{N,M_1,M_2} = \left[ \oint_1 + \oint_{1/\alpha}
\right]\frac{dy}{2\pi} \frac{y^N}{(y-1)^{M_1}(\alpha y-1)^{M_2}},~~~~~ 
\en
\eq
F^b_{N,M_1,M_2} = \left[ \oint_1 + \oint_{1/\alpha}
\right]\frac{dy}{2\pi} \frac{y^N}{(y-1)^{M_1}(1-by)^{M_2}}
\en 
(notice that the contours of integration are the same for the two
integrals)
\eq
v_2 = M_2 \frac{F^\alpha_{N-2, M_1, M_2}F^b_{N,M_1+1,M_2}- F^\alpha_{N, M_1+1,
  M_2}F^b_{N-2,M_1,M_2} }{F^\alpha_{N-1, M_1, M_2}F^b_{N,M_1+1,M_2}- F^\alpha_{N, M_1+1,
  M_2}F^b_{N-1,M_1,M_2}}. 
\en
When $\alpha = 1$ and $b=0$ we can compute both $F^\alpha$ and $F^b$
exactly \footnote{As explained in \cite{derr-evans} when $\alpha = 1$ one has to
take a single contour integral around $1$.}
\eq
F^{\alpha=1}_{N, M_1, M_2} =
\binom{N}{M_1+M_2-1},~~~~~~~~~~~~~F^{b=0}_{N, M_1, M_2} =
\binom{N}{M_1-1}~ .
\en
And for the velocity we get
\eq
v_2 = M_2 \frac{N-M_1 -2M_2}{N-1}.
\en
\sk

\sk

\section{Large N limit of the velocity}

We want now to consider the limit $N\rightarrow \infty$ with non-zero
densities of particles of both species $\rho_1 = M_1/N$ and $\rho_2 =
M_2/N$. To find the asymptotic formula for the velocity we have simply to
determine the asymptotic of $F^\alpha$ and $F^b$, which are easily given by
the steepest descent method. Both integrals have two saddle points which
correspond to the solutions of the equation
\eq
\frac{1}{y} - \frac{\rho_1}{y-1} - \frac{\rho_2 \kappa}{\kappa y-1}
\en  
where $\kappa = \alpha$ for $F^\alpha$, and  $\kappa = b$ while  for $F^b$.
The expression for the saddle points is
\eq\label{ys}
y_\kappa^\pm = \frac{\kappa +1 - \rho_1 -\kappa \rho_2 \pm \sqrt{(\kappa +1
    - \rho_1 -\kappa \rho_2)^2 -4\kappa (1-\rho_1
    -\rho_2)}}{2\kappa(1-\rho_1-\rho_2)} .
\en
It is easy to realize that both saddle points are on the real line, one is
situated between $1$ and $1/\kappa$, the other is situated to the right of
$1/\kappa$. 

When we compute $F^\alpha$ we can merge the contour around $1$ with
the one around $1/\alpha$ and let the resulting contour pass through
$y_\alpha^+$, hence $F^\alpha$ is dominated by the contribution from
$y_\alpha^+$. In computing $F^b$ the integral around $1/\alpha$ gives no
contribution because the integrand is holomorphic there. The contour
integral around $1$ can now be deformed only to pass through $y_b^-$,
because of the singularity present in $1/b$,  hence $F^b$ is dominated by
the contribution from  $y_b^-$.


In conclusion we get
\eq\label{velocity2}
v_2 = M_2\left(\frac{1}{y_\alpha^+} + \frac{1}{y_b^-} -1\right).
\en
By the same procedure one can find the total velocity of the particles of
kind $1$
\eq\label{velocity1}
v_1 = \frac{N}{y_\alpha^+ y_b^-}-(N-M_1)\left(\frac{1}{y_\alpha^+} + \frac{1}{y_b^-} -1\right).
\en
From the eq.(\ref{ys}) we see that the non-analyticities of the total velocities
(\ref{velocity2}, \ref{velocity1}) are located at the zeros of the square root, i.e. at
\eq\label{sing1}
(\alpha +1
    - \rho_1 -\alpha \rho_2)^2 -4\alpha (1-\rho_1
    -\rho_2) =0
\en
and at 
\eq\label{sing2}
(b +1
    - \rho_1 -b \rho_2)^2 -4b (1-\rho_1
    -\rho_2) =0.
\en
Eq.(\ref{sing1}) has solutions: 
\begin{itemize}
\item
for $\alpha<1$:  $\rho_2 =0$, $\rho_1=1-\alpha$;
\item
for $\alpha>1$:  $\rho_1 =0$, $\rho_2=\frac{\alpha-1}{\alpha}$.
\end{itemize}  
Eq.(\ref{sing2}) has solution only for $\rho_2 =0$ and $\rho_1 = 1-b$. This
means that if the densities of the particles are non zero there is no
phase transition.

\section{Conclusions}

In this paper we have studied an ASEP with two species of particles and
different hopping rates. We have formulated the computation of the cumulants of
the currents as an eigenvalue equation, and we have shown that this leads to
an integrable (\emph{\`a la} Yang-Baxter) transition matrix. This has
allowed us to employ the formalism of the Algebraic Bethe Ansatz to solve
the problem, by finding the Bethe Equations for an arbitrary number of
particles of each specie. The analysis of the Bethe equations gives in
principle all the cumulants of the currents. We found the exact formula for 
the velocity of the particles of type $2$, and computed its limit when the
size of the system goes to infinity, keeping non-zero densities for the
particles. We find this way that, when the densities are different from
$zero$, there is no phase
transition.

Our work can be extended in different directions. First, we think  
it would be interesting to use the Bethe Equation we found, to compute the
spectral gap  as a function of the hopping parameters $\alpha$ and $\beta$.  
We have briefly discussed in the Appendix the extension of the problem to a
larger number of species and different hopping rates, it would be also  nice to
work out the average velocity of particles of a given type as functions of
the hopping parameters. 
Another interesting possibility is to consider the problem on a lattice with
open ends and letting particles flow in and out of the system. We plan to
come back to these issues soon.

\section*{Acknowledgments}

It is a pleasure to thank Kirone Mallick and Sylvain Prolhac for useful
discussions and comments. 
This work has been supported by the ANR program ``GIMP'' ANR-05-BLAN-0029-01.

\appendix
\section{Algebraic Bethe Ansatz for the ASEP with one specie}\label{ABA1}


Let us consider the ASEP consisting of $m$ particles and $r$ holes on a ring of size
$\tilde N = p+r$. 
Each particle can jump into a neighbor site only if the site is
empty. The probability for jumping forward is 
$q\cdot dt$, while the probability for jumping backward is $p\cdot dt$.
Following \cite{der-lebo1} we consider the total distance covered
by all the particles in a time 
$t$, denoted by $Y_t$. In order to determine the
behavior of $Y_t$ we 
look at the joint probability $P_t(\cC,Y)$ of 
being at time $t$ 
in a configuration $\cC$ and having all the particles covered a total
distance $Y_t=Y$. 
The
generating function  
$$F_t(\cC) =
\sum_{Y=0}^\infty e^{\nu_{12} Y}P_t(\cC,Y), $$ which satisfies 
the following evolution equation
\eq
\frac{d}{dt}F_t(\cC) = \sum_{\cC'} [M_0(\cC, \cC') +  e^{\nu_{12}}  M_1(\cC, \cC') +
 e^{-\nu_{12}} M_{-1}(\cC, \cC')]F_t(\cC').
\en
where $M_1(\cC, \cC')dt$ is the transition probability for going from
the configuration $\cC'$ to the configuration $\cC$ \emph{and} moving a particle
\emph{forward} of one step, while $M_{-1}(\cC, \cC')dt$ correspond to a
particle moving \emph{backward} of one step and $M_0$ is the diagonal
part.
The
large time behavior of $\<< e^{\nu_{12} Y_t}\>>$ is determined by the largest
eigenvalue  $\lambda(\nu_{12})$ of the matrix transition matrix  $M_0(\cC,
\cC') +  e^{\nu_{12}}  M_1(\cC, \cC') + 
 e^{-\nu_{12}} M_{-1}(\cC, \cC')$.

\sk

The transition matrix $M_0(\cC, \cC') +  e^{\nu_{12}}  M_1(\cC, \cC') +
e^{-\nu_{12}} M_{-1}(\cC, \cC')$ can be diagonalized by means of  the Algebraic
Bethe Ansatz \cite{faddeev, golinelli-mallick1}. 
In our case we are led to consider
\eq
R_{a,b}(x,y) = 1 + \lambda(x,y) E
\en
where 
$$\lambda(x,y) = \frac{e^{\frac{x-y}{2}}-e^{\frac{y-x}{2}}}{p~
  e^{\frac{x-y}{2}}- q~e^{\frac{y-x}{2}}}.$$ 
The matrix $R_{a,b}(x,y)$ acts on $V_a\otimes V_b \rightarrow V_a\otimes
V_b$, where $V= \bC^2$, and in the basis $|00\>>, |01\>>, |10\>>, |11\>>$
the matrix $E$ reads
\eq\label{R1}
E =\left(
\begin{array}{cccc}
0&0&0&0\\
0&-q&p e^{-\nu_{12}}&0\\
0&q e^{\nu_{12}}&-p&0\\
0&0&0&0
\end{array}
\right)
\en
We define the matrices $L_{a,a_i}(x,y_i) = P_{a,a_i} R_{a,a_i}(x,y_i)$,
where $P_{a,b}$ is the permutation operator. 
We introduce also a matrix $\Omega$ which acts only on the auxiliary space,
in a diagonal way 
\begin{displaymath}
\Omega(\vec y) = 
 \left(
\begin{array}{cc}
\omega_1(\vec y)&0\\
0&\omega_2(\vec y)
\end{array}
\right)
\end{displaymath}
and its entries can depend on some auxiliary parameters $\vec y$.  
The reason for considering such a generalization of the problem we
started from, which in facts correspond to $\Omega$ equal to the
identity, comes from the ASEP with two species, as seen in the text.
For a system of size $\tilde N$ the monodromy matrix is constructed
by means of $L_{a,a_i}(x,\tilde y_i)$ \footnote{Notice that $y$s and $\tilde
 y $ can and will be in general different quantities.}
$$
\sT_{a\otimes \sH}(x,\vec {\tilde y}) = L_{a,a_{\tilde N}}(x,\tilde y_{\tilde
  N})\dots  L_{a,a_1}(x,\tilde y_1).  
$$
The transfer matrix is given  by 
$$
T(x,\vec y, \vec {\tilde y}) = \tr_a \left[\Omega(\vec y)\sT_{a\otimes
    \sH}(\vec {\tilde y})\right]. 
$$
The fact that
$[\Omega(\vec y)\otimes\Omega(\vec y), R(x,x')] = 0$, combined with the
Yang-Baxter equation,  implies that
the transfer matrices with different parameters $x$ and $x'$ commute
among themselves.
The transition matrix of the ASEP is obtained as the logarithmic derivative of
the transfer matrix in zero (at $\Omega = \rm{Id}$ and $y_i=0$).

Let us write the monodromy matrix   in the auxiliary
space as 
\eq
\sT (x, \vec {\tilde y}) =\left(
\begin{array}{cc}
A(x, \vec {\tilde y}) & B(x,\vec {\tilde y})\\
C(x, \vec {\tilde y}) & D(x, \vec {\tilde y})
\end{array}
\right),
\en
the Algebraic Bethe Ansatz proceeds by constructing an eigenvector acting
with  $B(\zeta_i, \vec{\tilde y})$ on a reference state. Our reference state is $|1\>>= 
\left( \begin{array}{c}
1\\
0
\end{array}  \right)
\otimes
\dots
\otimes
\left( \begin{array}{c}
1\\
0
\end{array}  \right)
$
corresponding to the completely full system, which is an eigenvector of the
transfer matrix. Indeed we notice that
$
A(x, \vec{\tilde y}) |1\>> = |1\>>, ~~D(x, \vec{\tilde y}) |1\>> = \prod_{k=1}^{\tilde N}\left(
  \frac{p\lambda(x,\tilde y_k)}{e^{\nu_{12}}}\right)|1\>>, ~~
C(x, \vec{\tilde y}) |1\>> = 0.
$  
Hence the eigenvalue of $|1\>>$ is 
$
\omega_1(\vec y) A(x, \vec{\tilde y}) + \omega_2(\vec y)
D(x, \vec{\tilde y})= \omega_1(\vec y) + \omega_2(\vec y)
\prod_{k=1}^{\tilde N}\left( p\lambda(x, \tilde y_k)\right).
$
We search now for eigenvectors of the form
\eq
|(\zeta_1,\dots, \zeta_r)\>> = B(\zeta_1)\dots B(\zeta_r) |1\>>
\en
and use the Yang-Baxter algebra, satisfied by the operators $A(x), B(x), C(x), D(x)$ as a
consequence of the Yang-Baxter equation 
\begin{eqnarray}\label{commut3}
[A(x), A(y)]&=&0,\\
A(x)B(z) &=& \frac{e^{\nu_{12}}}{p \lambda(z,x)}B(z) A(x) -
\frac{e^{\nu_{12}}(1-p \lambda(z,x))}{p \lambda(z,x)} B(x)A(z) \\
B(z) B(x) &=& B(x) B(z) \\
D(x) B(z) & = & \frac{e^{\nu_{12}}}{p\lambda(x,z)}B(z) D(x) -\frac{e^{\nu_{12}}(1- q \lambda(x,z))}{p
  \lambda(x,z)}B(x) D(z).  
\end{eqnarray}

The requirement $|(\zeta_1,\dots, \zeta_r)\>>$ to be an eigenvector can be
expressed in terms of a Bethe equation 
\eq
\frac{\omega_2(\vec y)}{\omega_1(\vec y)}\prod_{i \neq j} \left( \frac{\lambda(\zeta_i,\zeta_j)}{
    \lambda(\zeta_j,\zeta_i)}\right) \prod_{k=1}^{\tilde N} (p
\lambda(\zeta_j, \tilde y_k))= e^{\nu_{12}\tilde N}, 
\en
which fixes the values of $\vec \zeta$. 
The eigenvalue of the transfer matrix is given by
\eq
\Lambda(x) = \omega_1(\vec y) \prod_{i=1}^r \left( \frac{e^{\nu_{12}}}{p
    \lambda(\zeta_i,x)}\right) + \omega_2(\vec y) \prod_{i=1}^r \left(
  \frac{e^{\nu_{12}}}{p \lambda(x, \zeta_i)}\right) 
\prod_{k=1}^{\tilde N} \left(\frac{p \lambda(x, \tilde y_k)}{e^{\nu_{12}}}\right).
\en

\sk

We consider now the limit
$p\rightarrow 0, q=1$. In order to do that without getting a singular limit we
have to change the spectral parameters into 
$\zeta_i \rightarrow \zeta_i-\frac{\log(p)}{2}$.
Then the Bethe equations turn into the form
\eq
\frac{\omega_2(\vec y)}{\omega_1(\vec y)}\prod_{i \neq j} \left(-
  \frac{p~e^{(\zeta_j-\zeta_i)/2}-q~e^{(\zeta_i-\zeta_j)/2}}
{p~e^{(\zeta_i-\zeta_j)/2}-q~e^{(\zeta_j-\zeta_i)/2}}
\right) \prod_{k=1}^{\tilde N} \left(\frac{e^{(\zeta_j -\tilde
      y_k)/2}-p~e^{(\tilde y_k-\zeta_j)/2}}
{e^{(\zeta_j -\tilde y_k)/2} - q e^{(\tilde y_k-\zeta_j)/2}}
\right)= e^{\nu_{12} \tilde N} 
\en 
and for $p\rightarrow 0$ and $q=1$ we get
\eq
\frac{\omega_2(\vec y)}{\omega_1(\vec y)}\prod_{i \neq j} \left(-
e^{\zeta_i-\zeta_j} 
\right) \prod_{k=1}^{\tilde N} \left(\frac{e^{-\tilde y_k}}
{e^{-\tilde y_k} - e^{-\zeta_j}}
\right)=e^{\nu_{12} \tilde N}  
\en
Defining $Z_j = e^{-\zeta_j}$ we arrive at
\eq\label{bethe-1sp}
\frac{\omega_2(\vec y)}{\omega_1(\vec y)}\prod_{i \neq j} \left(-
\frac{Z_j}{Z_i}
\right) \prod_{k=1}^{\tilde N} \left(\frac{e^{- \tilde y_k}}
{e^{- \tilde y_k}- Z_j}
\right)=e^{\nu_{12} \tilde N}
\en
and the eigenvalue can be simply  written as
\eq\label{eigen1}
\Lambda(x) = e^{\nu_{12} r} \omega_1(\vec y) \prod_{i=1}^r \left(1-e^{x}Z_i \right)
\en

\section{Yang-Baxter equation for multi-species ASEP with different
  rates}\label{multi}

In this appendix we want to discuss to what extent the Baxterized form
of the $R-$matrix (\ref{baxterized}) can be generalized, in order to
describe a process with a number of species greater than $3$, and  a
hierarchical structure. Labelling the species with numbers from 1
to $n$, the hierarchy means that a  particle of kind $i$ can hop to
the right of a particle of 
kind $j$ only
if $i<j$. It is well known that if all these elementary processes
happen with the same rate, then  the $R-$matrix is simply given by the
Baxterization of the Hecke algebra \cite{rittenberg}. What we want to
consider here is 
the case when the hoppings among the particles depend on the species
involved in the hopping
\begin{center}

$ij \rightarrow ji$ ~~~with rate ~~$a_{ij}$~~~ if~~~ $i<j$.

\end{center}
Writing the matrix describing the hopping $ij \rightarrow ji$ as
$E^{(ij)}_{\alpha \beta, \gamma \sigma} = \delta_{\alpha
  i}\delta_{\beta j}(e^{\nu_{ij}} \delta_{\alpha
  \sigma}\delta_{\beta\gamma}- \delta_{\alpha
  \gamma}\delta_{\beta\sigma})$, we would
like to find an $R-$matrix of the form 
\eq\label{Rgen}
R(x,y) = 1 + \sum_{\{ij\}} g_{ij}(x,y)E^{(ij)},
\en
which satisfies the Yang-Baxter equation
\eq\label{YBE3}
R_{a,b}(y,z) R_{b,c}(x,z) R_{a,b}(x,y) = R_{b,c}(x,y) R_{a,b}(x,z)
R_{b,c}(y,z),
\en
and such that
\eq
\frac{d}{dx}R(x,y)\Big|_{x=y=0} = \sum_{\{ij\}} a_{ij} E^{(ij)}.
\en
Let us define for $i< j < k$ the projectors $P_{i,j}$ and
$P_{i,j,k}$, which act on $\bC^n \otimes \bC^n \otimes \bC^n$. $P_{i,j}$
projects on the states occupied only by particles of type $i$ and $j$, while
$P_{i,j,k}$ projects on states occupied only by particles $i$, $j$ and $k$.  
If we intertwine eq.(\ref{YBE3}) with $P_{i,j}$ we see that we reduce
to the problem with two types of particles (or equivalently one type
of particles and the empty sites), and we recover easily that $g_{ij}$
must be of the form
\eq
g_{i,j}(x,y) = 1-\frac{f_{i,j}(x)}{f_{i,j}(y)}.
\en
If we intertwine  eq.(\ref{YBE3}) with $P_{i,j,k}$ we recover the
problem with three types of particles treated in the main body of this
paper, and the Yang-Baxter equation implies that
\eq\label{relations}
f_{i,k}(x) = f_{i,j}(x) + b_{i,k}^{i,j};~~~~~~~f_{j,k}^{-1}(x) =
f_{i,k}^{-1}(x) + b_{j,k}^{i,k}.
\en
This means that all the functions $f_{i,j}(x)$ are determined in terms of a
reference one, which we chose to be $f_{1,2}(x)$, and of the parameters
$ b_{i,k}^{i,j}$ and $b_{j,k}^{i,k}$. Actually the relations in
(\ref{relations}) put also constraints on the $b$s. Indeed it is easy
to see that we must have
$$
b_{i,k}^{i,j} = \sum_{l=j}^{k-1} b_{i,l+1}^{i,l};~~~~~~~~
b_{j,k}^{i,k} = \sum_{l=i}^{j-1} b_{l+1,k}^{l,k}.
$$
Moreover if $i>1$ one can get $f_{i,j+1}(x)$ starting from$f_{i,j}(x)$
in two different ways
$$
f_{i,j}\rightarrow f_{i,j+1}~~~~~ \textrm{ or}~~~~~f_{i,j}\rightarrow
f_{i-1,j} \rightarrow f_{i-1,j+1} \rightarrow f_{i,j+1}. 
$$  
The previous relation fixes 
$$b_{i,j+1}^{i-1,j+1}=
\frac{b_{i,j}^{i-1,j}}{1-b_{i-1,j+1}^{i-1,j}b_{i,j}^{i-1,j}}~~~~~
     \textrm{ and}~~~~~ b_{i,j+1}^{i,j}= 
\frac{b_{i-1,j+1}^{i-1,j}}{1-b_{i-1,j+1}^{i-1,j}b_{i,j}^{i-1,j}}~~.$$

In conclusion we can chose as free parameters $b_{1,j+1}^{1,j}$ and
$b_{l-1,l}^{l-2,l}$, which in a problem with $n$ species are in number
of $2(n-1)$. Hence among the $n(n-1)/2$ rates $a_{ij}$, only $2(n-1)$
are independent, given the form of the $R-$matrix (\ref{Rgen}). 
This of course does not rule out the possibility that
the problem with generic rates is integrable, but one should look for
a more general $R-$matrix to prove it.

\bibliographystyle{amsplain}

\end{document}